\documentclass[useAMS,usenatbib]{mn2e}

\def\apj{{ApJ}}                 
\def\apjl{{ApJ}}                
\def\mnras{{MNRAS}}             
\def\nat{{Nature}}              

\usepackage{epsf}

\newcommand{\mpc}{\rm {h^{-1}Mpc }}
\newcommand{\ud}{\textrm{d}}

\title{Non-Perturbative Effects of Geometry in Wide-Angle Redshift Distortions}

\author[]{P\'eter P\'apai$^{12}$, Istv\'an Szapudi$^{1}$\\ 
$^1$Institute for Astronomy, 
University of Hawaii, 2680 
Woodlawn Dr, Honolulu, HI 96822\\ $^2$Department of Physics and Astronomy, 
University of Hawaii at Manoa, 2505 Correa Rd, Honolulu, HI 96822
}

\begin{document}

\maketitle

\begin{abstract}

We use the formalism of \cite{Szapudi2004}
to derive full explicit expressions for the 
linear two-point correlation function, including redshift space distortions
and large angle effects.
We take into account a non-perturbative geometric term in 
the Jacobian, which is still linear in terms of the dynamics.
This term had been identified previously 
\citep{Kaiser1987,HamiltonCulhane1996}, but has been 
neglected in all subsequent explicit calculations
of the linear redshift space two-point correlation function.
Our results represent a significant
correction to previous explicit expressions and are in excellent
agreement with our measurements in the Hubble volume simulation.

\end{abstract}

\section{INTRODUCTION}

Redshift distortions represent a curse disguised as a blessing for 
high precision cosmological applications.
Radial coordinates of redshift surveys contain limited phase space
information, which in principle can be used to constrain theories
more than configuration information alone; moreover, velocities are
sensitive to structure outside of the survey boundaries which potentially
translates into a larger ``effective volume''. On the other hand, redshift
distortions are plagued with non-linearities, both on large and small scales,
therefore in worst case they could amount to poorly understood
contamination of the configuration space data. Our aim is to extend
the theory of linear redshift distortions such that large angle information
could be successfully extracted from galaxy surveys.

The work of \cite{DavisPeebles1983} and \cite{Peebles1980} showing
that redshift distortions affect the power spectrum spawned a lot
of activity. The all-important
linear, plane-parallel limit was first calculated by 
\cite{Kaiser1987}, showing that the effect on the power spectrum
corresponds to ``squashing''. The other well known ``fingers of God''
effect dominates small scales, and is irrelevant for our present
study. The Kaiser formula has been generalized for real space soon
after \citep[e.g.,][]{Hamilton1993,ColeEtal1995}. These theories have been used to analyze 
surveys such as the Point Source Catalog Redshift (PSCz) Survey
\citep{TadrosEtal1999}, the Two-Degree Field Galaxy Redshift
Survey \citep[2dFGRS;][]{PeacockEtal2001,HawkinsEtal2003,TegmarkEtal2002},
and the Sloan Digital Sky Survey
\citep[SDSS;][]{ZehaviEtal2002}.

The distant observer approximation only holds if pairs are separated
by a small angle. This means that a large fraction of pairs needs to be
thrown away from modern wide angle redshift surveys when they are
analyzed in this limit. These pairs are typically fewer and noisier
than close pairs, but if our aim is to extract as much information
as possible from a given survey, it would be desirable to add them in.
\cite{HamiltonCulhane1996} related the 
``$\omega$-transform'', a complexified Mellin-like transform, of the
two-point correlation function to that of redshifted $\omega$-space
correlation function.
The resulting spherical $\omega\ell m$ expansion is approximately orthogonal
to redshift space distortions. This expansion truncated at an appropriate
mode was used in \cite{TegmarkEtal2002} to analyze data in this 
transform space.
The first explicit perturbation
theory calculation in coordinate space was performed in
\cite{SzalayEtal1998}. The result is a simple-to-use
finite expression, but only in coordinate space: in Fourier
space an infinite series will result for the redshift distorted
analog of the power spectrum. These formulae were later further 
generalized to include high-z effects in various cosmologies
by \cite{MatsubaraEtal2004}. These calculations provide essential
input for the pixel based Karhunen-Lo\'eve (KL) or quadratic
likelihood analyses \citep[e.g.,][]{VogeleySzalay1996,
TegmarkEtal1997}, since distant observer approximation
is not valid for modern wide angle galaxy surveys.
The theory has been applied in  several 
subsequent analyses of wide angle redshift
surveys, such as \cite{PopeEtal2004} and \cite{OkumuraEtal2008}.
Despite its elegance, the theory did not agree well with dark matter
simulations. \cite{Scoccimarro2004} pointed out that this might be
due to non-perturbative effects.

\cite{Szapudi2004} reanalyzed the redshift distortion problem from
group theoretical point of view showing that tripolar spherical harmonics
provide an excellent basis for expansion, and result in especially
compact formulae. In addition it provided specific coordinate systems,
one of which recovers the Legendre expansion of \cite{SzalayEtal1998},
while the other represents the same information in an even simpler 
two-dimensional Fourier mode expansion. 
We use this formalism to take into account
a term in the Jacobian, previously neglected in all explicit calculations,
to derive the full linear redshift distorted correlation function. 

Kaiser's original work starts with
the full linear Jacobian. It contains a term
negligible for small angles, that is
linear in terms of the small fluctuations, and is essentially non-linear
from the point of view of geometry: it contains a $1/r$ prefactor. 
Moreover, this term,
if expanded in bipolar spherical harmonics (or any other way),
would contribute infinite coefficients. 
Because of the presumed subdominance due
to the prefactor, and complexity of the calculation, this
term was neglected in all previous coordinate space expressions, 
although it is represented in the $\omega$-space expansion of 
\cite{HamiltonCulhane1996}. In this paper we introduce a hybrid
approach, where we leave the essentially non-perturbative terms in
the expansion intact; our tripolar expansion coefficients will still
contain angular variables in a specific way. 
As we show later, this hybrid procedure results
in a finite number of terms, and it provides significant corrections
and improvement in the agreement with simulations. In retrospect, 
the omission of this term, while intuitively reasonable, is not justified,
as its contribution can become important on the most interesting scales
of tens of $\mpc$'s.

In the next \S 2 we present the theory of linear redshift distortions
including results from the geometric term in the Jacobian. 
We follow closely the formalism of \cite{Szapudi2004}, mainly focusing
on the new aspects of this calculation. For reference, we print the
full result, which has about twice as many terms as previously.
In \S 3 we compare our results with preliminary measurements 
in the Hubble volume simulations, and present our conclusions.

\section{REDSHIFT DISTORTION OF THE TWO-POINT CORRELATION FUNCTION }

We use linear perturbation theory to predict the redshift distorted
two-point correlation function in terms of the underlying power
spectrum. Our calculation is based directly on the tripolar expansion
formalism of  \cite{Szapudi2004}, therefore our focus will be on the
additional terms arising from the Jacobian. 

The exact mapping between real and redshift space is
 $s_{i}=x_{i}-fv_{j}\hat{x}_{i}\hat{x}_{j}$, where
the ''hat'' denotes the proper unit vector, $f=\frac{\Omega^{0.6}}{b}$
and the velocity has units which provides that its divergence is equal
to the density up to linear order. From this, one can calculate the derivative
of this matrix: $\partial s_i/\partial x_k = \delta_{ik}+O_{ik}(v)$ where $O$ is linear in $v$.
This results in a linear Jacobian $J=1+TrO = 1-f\hat{x}_{i}\hat{x}_{j}\partial_i
v_{j}-2f\frac{x_{j}v_{j}}{x^{2}}$. The last term in the previous expression
is usually omitted due to the fact that it scales with $1/x$, i.e. it would
tend to zero for large distances, which loosely correspond to large angles
as well. Closer examination of this term shows that it is of the same
order as the previous term, not only in perturbation expansion 
(linear), but also in order of magnitude. Our goal is to propagate
this new term through the full calculation.

The linear density contrast and the
two-point function can be expressed in the usual fashion.
\begin{eqnarray}
	\delta_{s}(x)= \int \frac{\ud^{3}k}{(2\pi)^{3}}e^{ik_{j}x_{j}}\left [ 1+f(\hat{x}_{j}\hat{k}_{j})^{2}-i2f\frac{\hat{x}_{j}\hat{k}_{j}}{xk}\right ] \delta (k) \label{dens}\\
	\langle\delta_{s}(x_{1})\delta^{*}_{s}(x_{2})\rangle = \int \frac{\ud^{3}k}{(2\pi)^{3}}P(k)e^{ik(x_{1}-x_{2})} \cr \left [ 1+\frac{f}{3}+\frac{2f}{3}P_{2}(\hat{x}_{1}\hat{k})-\frac{i2f}{x_{1}k}P_{1}(\hat{x}_{1}\hat{k})\right ] \cr \left [ 1+\frac{f}{3}+\frac{2f}{3}P_{2}(\hat{x}_{2}\hat{k})+\frac{i2f}{x_{2}k}P_{1}(\hat{x}_{2}\hat{k})\right ], \label{2pt}
\end{eqnarray}
where $P_{1}$ and $P_{2}$ are Legendre polynomials and $P(k)$ is the
linear power spectrum. The third term in each of the brackets correspond
to the extension of the previous results; these would tend to zero
in the plane parallel limit. At
wide angles, the separation between the galaxies and the
distance between a galaxy and the observer are of the same order, therefore
$kx$ is of order unity. This shows explicitly that the order of this
term can be as large as the previous, and the detailed calculation confirms
this. 

Next we express the angular dependence of the correlation function with
tripolar spherical harmonics.
\begin{eqnarray}
S_{l_1 l_2 l}(\hat x_1,\hat x_2, \hat x) \cr
\equiv \sum_{m_1,m_2,m}
\left( \begin{array} {ccc} l_1 & l_2 & l \\ m_1 & m_2 & m
             \end{array} \right) 
      C_{l_1 m_1}(\hat x_1) C_{l_2 m_2}(\hat x_2) C_{l m}(\hat x)
\end{eqnarray}
We use $x$ for denoting $x_1 - x_2$. On the right hand side one can
find the Wigner $3j$ symbols and we define the
normalized spherical functions as $C_{lm} =
\sqrt{4\pi/2l+1}Y_{lm}$; these latter result in simpler expressions.

Eq.~(\ref{2pt}) has become more complex with the additions,
$x_1$ and $x_2$ appear in the
denominator resulting in the following angular dependence
\begin{eqnarray}
	x_{1} = g_{1} x = \frac{\sin(\phi _2)}{\sin(\phi _2 - \phi _1)} x \\
	x_{2} = g_{2} x = \frac{\sin(\phi _1)}{\sin(\phi _2 - \phi _1)} x.
\end{eqnarray}
Expanding these terms into tripolar spherical harmonics would yield infinite
terms, but simplification arises from the fact that they can be factored
out of the integrals. All the rest can be expanded as in \cite{Szapudi2004},
resulting in finite expressions. We introduce
$\phi _1$ to denote the angle between $\hat x_1$ and $\hat x$ and
$\phi _2$ for the angle between $\hat x_2$ and $\hat x$. We emphasize that
the coefficients of this (quasi-)tripolar expansion still has an angular
dependence in the form of $g_{1}$ and $g_{2}$:
\begin{eqnarray}
	\xi{}_s =  \sum_{l_1 l_2 l} B^{l_1 l_2 l} (x,\phi _1,\phi _2)S_{l_1 l_2 l}(\hat x_1,\hat x_2, \hat x).
\end{eqnarray}
After performing the expansions, only a finite number of coefficients
survive. For reference, 
the ones from \cite{Szapudi2004} are:  
\begin{eqnarray}
  B^{000}(x) = (1+\frac{1}{3}f)^2 \xi^2 _0(x)\cr
  B^{220}(x) =\frac{4}{9\sqrt{5}}f^2 \xi^2 _0(x)\cr
  B^{022}(x)= B^{202}(x) = 
  -(\frac{2}{3}f+\frac{2}{9}f^2)\sqrt{5} \xi^2 _2(x)\cr 
  B^{222}(x) = \frac{4\sqrt{10}}{9\sqrt{7}}f^2\xi^2 _2(x)\cr
  B^{224}(x) = \frac{4\sqrt{2}}{\sqrt{35}}f^2\xi^2 _4(x);
\end{eqnarray}\\
and the new terms, the main result of this paper, are
\begin{eqnarray}
	B^{101}(x,\phi _1,\phi _2) = -(2f+\frac{2}{3}f^2)\frac{\sqrt{3}}{g_1 x} \xi_{1}^{1}(x) \cr
	B^{011}(x,\phi _1,\phi _2) = (2f+\frac{2}{3}f^2)\frac{\sqrt{3}}{g_2 x} \xi_{1}^{1}(x)\cr
	B^{121}(x,\phi _1,\phi _2) = \frac{4\sqrt{2}}{\sqrt{15}}f^2\frac{1}{g_1 x} \xi_{1}^{1}(x) \cr
	B^{211}(x,\phi _1,\phi _2) = - \frac{4\sqrt{2}}{\sqrt{15}}f^2\frac{1}{g_2 x} \xi_{1}^{1}(x)\cr
	B^{123}(x,\phi _1,\phi _2) = \frac{4\sqrt{7}f^2}{\sqrt{15}g_1 x} \xi_{3}^{1}(x) \cr
	B^{213}(x,\phi _1,\phi _2) = - \frac{4\sqrt{7}f^2}{\sqrt{15}g_2 x} \xi_{3}^{1}(x)\cr
	B^{110}(x,\phi _1,\phi _2) = -\frac{4f^2}{\sqrt{3}g_1 g_2 x^2}\xi_{0}^{0}(x) \cr
	B^{112}(x,\phi _1,\phi _2) = -\frac{4\sqrt{10} f^2}{\sqrt{3}g_1 g_2 x^2}\xi_{2}^{0}(x),
\end{eqnarray}
where $\xi_{l}^{m}(x) = \int \ud k/2\pi ^2 k^m j_l(xk) P(k)$ with $j$ being the spherical Bessel function.

\begin{figure*}
    \begin{minipage}{175mm}
    \begin{center}
      \leavevmode
      \epsfxsize=\columnwidth    
      \epsfbox{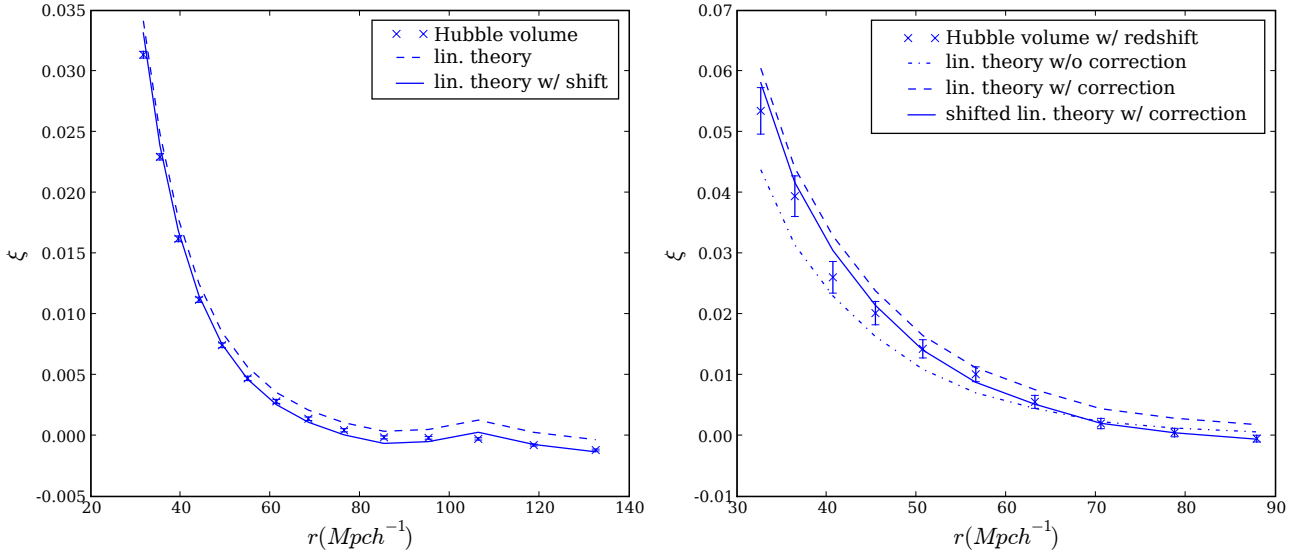}
    \end{center}
\caption{(Left) The measurement of the correlation function without
redshift distortion of the Hubble volume simulation (symbols) compared
with linear theory(dashed and solid lines). The error bars were estimated from $9^3$
subvolumes of th Hubble volume. Shifting the theory by 0.00081 downward,
motivated by the integral constraint,
provides an excellent fit to the data. 
(Right) Redshift
distorted correlation function of the Hubble volume simulation
(symbols) at constant opening angle (0.71 radian) and while the ratio of
the distances of the particles in the pair are kept fixed (at
1.57). The error bars were estimated as before. The lines indicate the linear theories
with (this paper) and without the geometric terms. The solid line is the corrected theory with a
downshift of 0.0016. The integral constraint correction is
expected to be larger since the average of
the two point function is larger.
}
    \label{fig}
    \end{minipage}
\end{figure*}

For further elaboration we choose coordinate system a) from
\cite{Szapudi2004}. This corresponds to our 
previous choice of angles with $\phi _1$, $\phi _2$, with which
the $S_{l_1 l_2 l}(\hat x_1,\hat x_2, \hat x) = S_{l_1 l_2
l}(\pi/2,\phi_1,\pi/2,\phi_2,\pi/2,0)$ functions can be expressed using
cosines and sines only. Using the same notation as \cite{Szapudi2004}:
\begin{eqnarray}
	\xi_s(\phi_1,\phi_2,x) = 
	 \sum_{n_1,n_2=0,1,2} a_{n_1 n_2} \cos(n_1 \phi_1) \cos(n_2 \phi_2) \cr + b_{n_1 n_2} \sin(n_1 \phi_1)\sin(n_2 \phi_2).
\end{eqnarray}
Again, for reference, the previously calculated coefficients are
\begin{eqnarray}
	a_{00 }=
\left( 1 + \frac{2f}{3} + \frac{2f^2}{15} \right)
     \xi^2 _0(x) - \cr \left( \frac{f}{3} +
     \frac{2f^2}{21} \right) \xi^2 _2(x)  + 
  \frac{3f^2}{140}\xi^2 _4(x)\cr
a_{02}=a_{20}= \left( \frac{-f}{2} - 
        \frac{3f^2}{14} \right) \xi^2 _2(x) + 
     \frac{{f}^2}{28}\xi^2 _4(x)  \cr
a_{22} =  \frac{{f}^2}{15}\xi^2 _0(x) - 
     \frac{f^2}{21}\xi^2 _2(x) + 
     \frac{19f^2}{140}\xi^2 _4(x) \cr
b_{22} = \frac{{f}^2}{15}\xi^2 _0(x) - 
     \frac{f^2}{21}\xi^2 _2(x) - 
     \frac{4f^2}{35}\xi^2 _4(x);	
\end{eqnarray}
and the new expressions of this work correspond to
\begin{eqnarray}
	a_{10}= \frac{\tilde{a}_{10}}{g_1} =(2f+\frac{4f^2}{5} )\frac{1}{g_1 x}\xi_1 ^1 -\frac{1}{5} \frac{f^2}{g_1 x}\xi_3 ^1  \cr
	a_{01}=\frac{\tilde{a}_{01}}{g_2} =-(2f+\frac{4f^2}{5} )\frac{1}{g_2 x}\xi_1 ^1 +\frac{1}{5} \frac{f^2}{g_2 x}\xi_3 ^1  \cr	
	a_{11}=\frac{\tilde{a}_{11}}{g_1 g_2} =\frac{4}{3}\frac{f^2 }{g_1 g_2 x^2 }\xi_0 ^0 -\frac{8}{3} \frac{f^2 }{g_1 g_2 x^2 }\xi_2 ^0 \cr
	a_{21}=\frac{\tilde{a}_{21}}{g_2} =-\frac{2}{5}\frac{f^2 }{g_2 x}\xi_1 ^1 +\frac{3}{5} \frac{f^2 }{g_2 x}\xi_3 ^1 \cr
	a_{12}=\frac{\tilde{a}_{12}}{g_1} =\frac{2}{5}\frac{f^2 }{g_1 x}\xi_1 ^1 -\frac{3}{5} \frac{f^2 }{g_1 x}\xi_3 ^1 \cr
	b_{11}=\frac{\tilde{b}_{11}}{g_1 g_2} =\frac{4}{3}\frac{f^2 }{g_1 g_2 x^2 }\xi_0 ^0 +\frac{4}{3} \frac{f^2 }{g_1 g_2 x^2 }\xi_2 ^0 \cr
	b_{21}=\frac{\tilde{b}_{21}}{g_2} =-\frac{2}{5}\frac{f^2 }{g_2 x}\xi_1 ^1 -\frac{2}{5} \frac{f^2 }{g_2 x}\xi_3 ^1 \cr
	b_{12}=\frac{\tilde{b}_{12}}{g_1} =\frac{2}{5}\frac{f^2 }{g_1 x}\xi_1 ^1 +\frac{2}{5} \frac{f^2 }{g_1 x}\xi_3 ^1.	
\end{eqnarray}

It is worth to emphasize again that the angular dependence 
$g_1$ and $g_2$ is suppressed for clarity in the above formulae,
but it is obviously carries through according to the definition
of these functions. If 
the equivalence of the configurations $(\phi _1, \phi _2 )
\rightarrow (\pi - \phi_2, \pi - \phi_1)$ is taken into account
(same pairs can be counted twice), the number of independent 
new coefficients is five, i.e. the number of terms approximately
doubled. Next we explore the relevance of these calculations,
and compare the theoretical predictions with measurements in dark
matter only $N$-body simulations.

\section{DISCUSSION AND SUMMARY}

To understand our results, we 
expanded our formulae to identify leading order corrections
to the Kaiser limit.

The leading order corrections to the distant observer approximation
are second order. Using the notation $\frac{1}{2}(\phi _1 +
\phi _2) = \phi$ and $\frac{1}{2}(\phi _2 - \phi _1) = \Delta \phi$,
and keeping leading order terms in $\Delta \phi$ results in
\begin{eqnarray}
	\xi_s(\phi,\Delta \phi,x) \cr = a_{00}+2a_{02}\cos (2\phi) + a_{22}\cos^2 (2\phi)+b_{22}\sin^2(2\phi) \cr
	+\bigg[-4a_{02}\cos (2\phi)-4a_{22}-4b_{22}\bigg] \Delta \phi^2 \cr
	+\bigg[-4\tilde{a}_{10}\cot^2 (\phi)+4\tilde{a}_{11}\cot^2 (\phi) \cr
	 -4\tilde{a}_{12}\cot^2 (\phi)\cos (2\phi)+4\tilde{b}_{11}-8\tilde{b}_{12}\cos^2 (\phi) \bigg] \Delta \phi^2 + \cr
	+O(\Delta \phi^4 ). 
	\label{geom}
\end{eqnarray}
The first line of eq. (\ref{geom}) corresponds to
the Kaiser formula $(\Delta \alpha
= 0)$.The next line contains leading order corrections corresponding
to previous work only, and the third line collects leading order corrections
from the geometric term in the Jacobian. These are all of the same order,
reassuring the need of keeping the geometric non-perturbative terms.
We conjecture that the terms containing the $\cot^2 (\phi)$ 
could be responsible for the reported failure of the linear theory for 
small angles along the line of sight \citep{OkumuraEtal2008}.

As a preliminary test of the validity of our calculations,
we measured correlation functions in the Hubble volume simulation
\citep{EvrardEtal2002},
using cosmological parameters $\sigma_8 = 0.9$, $n_s = 1$, $\Omega_{m}=0.3$, 
$\Omega_{\lambda}=0.7$, $h = 0.7$, $\Omega_b h^2 = 0.0196$ and a
volume of $(3000Mpch^{-1})^3$,
with and without redshift distortions. The volume of
the simulation was divided into $9^3$ subvolumes to 
obtain the error bars.

The left panel of the figure shows the
measured and the theoretical two-point functions without redshift 
distortions. The theory agrees with the measurements only after a shift
by a constant. This is due to the ``integral constraint'' problem
\citep[e.g.,][]{Peebles1980}, possibly compounded with slight non-linear
effects. This constant represents a bias which is approximately equal to the
average of the two-point correlation function over the survey area. It can
be determined several ways (see discussion below).

Next, an observer was placed at the
center of each subvolume and the mapping between real and redshift
space was performed using the velocities recorded in the simulation.
The correlation function was then measured using brute force counting
of pairs in high resolution bins matching our choice of coordinate system
described earlier.
The right panel of the figure presents wide angle redshift distortion
theory both with and without non-perturbative geometric corrections.
The latter cannot be made to agree with the measurements
even using a constant offset due to the integral constraint. 
In contrast, the
theory presented in this paper provides
excellent agreement with the measurements if the effects of integral
constraint are taken into account. Note that this shift corresponding
to the latter is expected to be larger with redshift distortions
included simply because the two-point function is enhanced on large scales.

While one can simply fit this constant shift, corresponding to throwing
away a constant from the two-point correlation function \cite{FisherEtal1993},
we have estimated it in two more ways: Monte Carlo integrating the
theoretical expression for the correlation function, and empirically measuring
the variance of the average density on the scales of the subsamples. All
three methods are consistent with each other; the figure uses the empirical
variance over subsamples. Note that in applications, the first
method, i.e. discarding a constant from the theory, 
is the most prudent procedure to follow, 
since fluctuations on the scale of the full survey are not measurable.

While these measurements are preliminary in the sense that we did not
try to span the full parameter space of wide angle redshift distortions,
the results presented in this figure appear to be typical:
any other configurations we measured showed similar improvement. Scanning
the full parameter space with our present brute force two-point correlation
function code would be impractical, since we need a very large
number of pairs in each bin to beat down the error bars enough that the 
difference
between the two theories can be reliably measured. Although
we developed a fast grid based code as well, we
found that at these small values of the correlation function the pixel 
window function effects become important. These are more complex
for the redshift distorted correlation function depending on three variables
than in real space. 
Such effect should be modeled very accurately before one could fully
span the available parameter space.

A few simple extensions and modifications of our theory are needed
for practical applications when measuring the two-point function 
\cite{OkumuraEtal2008}, or when using our results to estimate a 
theoretical covariance matrix for a Karhunen-Lo\'eve (KL) analysis, 
\citep[see][for details]{PopeEtal2004}. If the sample is not volume
limited,  the redshift space density contrast is defined through the 
redshift space selection function ($\Phi(r)$). 
The effect of this can be taken into account by $g \rightarrow 2g/\alpha$.
Where $\alpha = \frac{\ud \log(r^2 \Phi(r))}{\ud \log(r)}$. The local bias 
can be neglected if we only deal with pairs further away from the observer
than the correlation length and motion of the local group 
can be transformed out by using the frame of the cosmic microwave
background. 
These problems have been discussed in detail by \cite{HamiltonCulhane1996},
and the solutions are exactly analogous in our case.

Note that the integral constraint problem does not appear in KL analysis
where only modes orthogonal to the average density are used. This is
more elegant than the simple treatment we have given here, but the essence
of it is the same: regarding the constant in the two-point correlation
function as a nuisance parameter accomplishes the same for direct
applications of our formulea.

With these caveats we conclude that our theory of wide angle redshift
distortions yielded simple-to-use explicit formulae, which
agree with simulations. The corrections to previous formulae represent
significant improvement at modest cost in complexity. 
Possible generalizations along
the lines of \cite{MatsubaraEtal2004} are left for future research. 

\section{ACKNOWLEDGMENT}

We thank Mark Neyrinck for  
carefully reading the manuscript.
This work was supported by NASA grant
NNG06GE71G and NSF grant AMS04-0434413.

\end{document}